\documentclass[%
aip,
jmp,%
amsmath,amssymb,
reprint,%
]{revtex4-1}

\usepackage{graphicx}
\usepackage{dcolumn}
\usepackage{bm}

\begin{document}

\preprint{APS/123-QED}


\title{Thermoelectric property of a one dimensional channel in the presence of a transverse magnetic field}

\author{Chengyu Yan}
\email{uceeya3@ucl.ac.uk}
\affiliation{
	London Centre for Nanotechnology, 17-19 Gordon Street, London WC1H 0AH, United Kingdom\\
}%
\affiliation{
	Department of Electronic and Electrical Engineering, University College London, Torrington Place, London WC1E 7JE, United Kingdom
}%
\affiliation{
	Micronova, Aalto University, Tietotie 3, Otaniemi, Espoo 02150, Finland
}%
\author{Michael Pepper}
\affiliation{
	London Centre for Nanotechnology, 17-19 Gordon Street, London WC1H 0AH, United Kingdom\\
}%
\affiliation{
	Department of Electronic and Electrical Engineering, University College London, Torrington Place, London WC1E 7JE, United Kingdom
}%
\author{Patrick See}
\affiliation{
	National Physical Laboratory, Hampton Road, Teddington, Middlesex TW11 0LW, United Kingdom\\
}%
\author{Ian Farrer}
\affiliation{%
	Department of Electronic and Electrical Engineering, University of Sheffield, Sheffield, S1 3JD, United Kingdom\\
}%
\author{David A. Ritchie}
\affiliation{%
	Cavendish Laboratory, J.J. Thomson Avenue, Cambridge CB3 OHE, United Kingdom\\
}%

\author{Jonathan Griffiths}
\affiliation{%
	Cavendish Laboratory, J.J. Thomson Avenue, Cambridge CB3 OHE, United Kingdom\\
}%

\date{\today}
             
\begin{abstract}
	
We studied the thermal conduction through a quantum point contact (QPC), defined in 		GaAs-${\mathrm{Al}}_{\mathrm{x}}$${\mathrm{Ga}}_{1\mathrm{\ensuremath{-}}\mathrm{x}}$As heterostructure, in the presence of a transverse magnetic field. A shift in the position of thermo-voltage peak is observed with increasing field. The position of the thermo-voltage peak follows the Cutler-Mott relation in the small field regime (B $<$ 0.5 T); it starts diverging from the Cutler-Mott relation in the moderate field regime, where a cubic magnetic field term dominates over the trivial quadratic term; eventually the shift saturates in the large field regime (B $>$ 3.0 T). Our results suggest that additional calibration is necessary when using QPC as thermometry, especially when the transverse magnetic field is applied.

\end{abstract}

\maketitle

Thermal and electric conduction, in a conducting system, are generally strongly coupled to each other, the mode that carries charge is the same one that carries energy (heat).  The coupling is found to be well established in two-dimensional electron gas (2DEG) in both the low magnetic field and integer quantum Hall regime\cite{PFC03,PFT04,ZCK09,GJJ09}, as well as one-dimensional electron gas\cite{STR89,MVB90,ANP00}. Some recent studies on the fractional quantum Hall regime (FQHE), where electron-electron interaction dominates, show different results regarding thermal conduction\cite{KFP94,BOI10,GG16,SGR17}. In all these works, quantum point contacts (QPCs) or quantum dots (QDs) have been widely employed to probe the local temperature and reveal the heat flow through the system. However, the role of QPCs or QDs in terms of heat flow has not been well characterized, especially in the presence of a transverse magnetic field.

In the present work, we focus on the impact of a transverse magnetic field on thermal conduction through a QPC. It is noticed the observations are well captured by Cutler-Mott relation at zero field; however, a noticeable discrepancy between Cutler-Mott relation and experimental results occurs at a finite magnetic field. A cubic magnetic field term, in terms of the position of the thermo-voltage peaks, dominates over the trivial quadratic term with a moderate magnetic field. Meanwhile, an anomalous lattice temperature dependence associated with the field is also revealed. Our results suggest that additional calibration is necessary when using QPC as thermometry.

\begin{figure}
	
	\includegraphics[height=4.2in,width=3.0in]{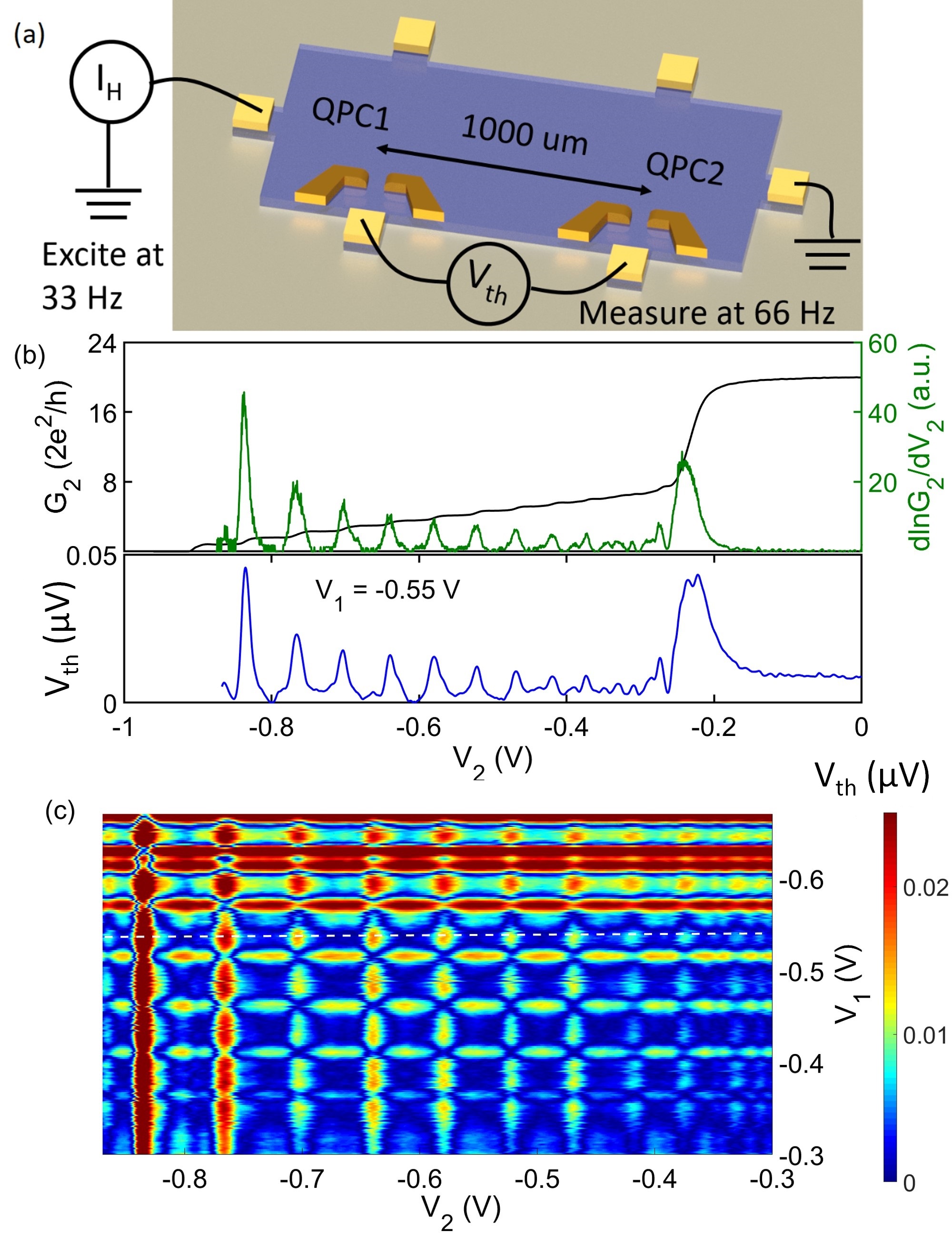}
	
	\caption{Experiment setup and thermo-voltage at zero magnetic field. (a) Schematic of experiment setup. The squares at the edge of the mesa are Ohmic contacts. The shining metallic gates define QPC1 and 2, and the QPCs are separated by $\sim$1000 $\mu$m. The large separation ensures the coupling between the two QPCs is negligible. The schematic is not to scale.
	(b) The upper panel shows conductance characteristic of QPC2. The lower panel illustrates representative results of V$_{th}$ as a function of V$_2$ with QPC1 set to 3G$_0$ (G$_0$ = $\frac{2e^2}{h}$). (c) V$_{th}$ spectrum with both QPC1 and QPC2 set to quasi-1D regime (V$_1$, V$_2$ $\leqslant$ -0.3 V). The white dashed cut corresponds to results in (b). }
	
	\label{fig:1}
	
\end{figure}

The thermoelectric property of a system is closely associated with a thermal gradient across the system. In the experiment setup [see note 1 of supplementary information for detailed setup] shown in  Fig. \ref{fig:1}(a) the whole system consists of three sections, once the split gates that define the quantum point contacts (QPCs) deplete electrons underneath the gates, namely the large 2DEG area with temperature $T_0$ and the relatively small area enclosed by QPC1 (QPC2) with temperature $T_1$ ($T_2$). The QPCs serve as thermal valve and local thermometry\cite{MVB90,ANS98,ANP00}. A heating current $I_H$ (at frequency $f_0$, 33 Hz; 1 $\mu$A) is fed to the 2DEG to increase $T_0$. Since $T_0 > T_{1,2}$, a thermo-voltage $V_{th}$ is built between the large and small area (measured at 2$f_0$, because heating power $P \propto I_H^2$). $V_{th}$ depends on the conductance of the QPCs, it vanishes if the QPCs are transparent (i.e. the QPCs are set to  $n \times \frac{2e^2}{h}$, $n$ is an integer), otherwise it takes a finite value. This behavior is well characterized by Cutler-Mott relation\cite{CM69,SI86} $$ \frac{V_{th,i}}{T_i - T_0}=-\frac{\pi^2 k_B}{3e}(T_i + T_0)\frac{\partial lnG_i}{\partial \mu_{i}} \eqno (1)$$ where $V_{th,i}$ is the thermo-voltage between $i$th QPC and common 2DEG, $k_B$ is the Boltzmann constant, $e$ is the elementary electron charge, $G_i$ is the electric conductance of $i$th QPC and $\mu_i$ is the chemical potential within the QPCs. It is necessary to comment that the usage of two QPCs minimizes the effect due to temperature fluctuation in the large 2DEG region\cite{ANS98,ANP00}. As a result, we focus on the net thermo-voltage drop between the two small area  $V_{th}$ = $V_{th,1}$ - $V_{th,2}$. However, most of the results can be obtained with a single-QPC setup.

Fig. \ref{fig:1} (b) demonstrates the excellent agreement between Cutler-Mott relation (upper panel) and measured $V_{th}$ (lower panel), which can be used to calibrate the system. The thermo-voltage was recorded by sweeping gate voltage applied to QPC2 whereas QPC1 was set to 3G$_0$. Tuning $V_1$ and $V_2$ (gate voltage applied to QPC1 and 2) independently, we could extract a detailed $V_{th}$ spectrum, as illustrated in Fig. \ref{fig:1}(c). Well organized patterns were observed. The bright horizontal segments happened when QPC2 was transparent, and QPC1 was not; vice versa for the vertical segments. The dark square region corresponded to the situation when both QPC1 and QPC2 were transparent. A complete spectrum, including the 2D regime, can be found in supplementary material, S1.

\begin{figure}
	
	\includegraphics[height=3.8in,width=3.2in]{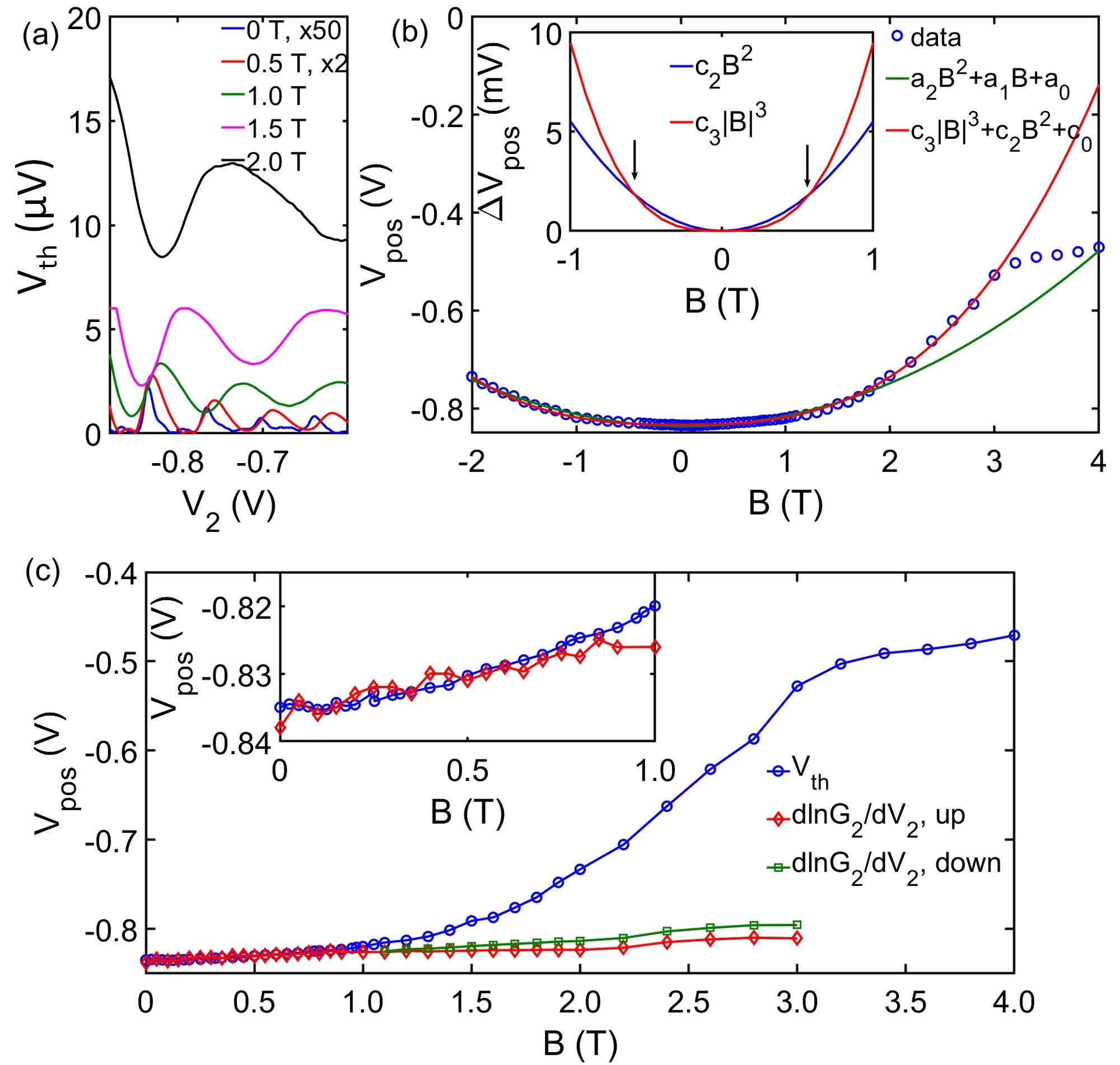}
	
	\caption{Magnetic field dependence of  V$_{th}$ peak position. (a) Trace in Fig. 1(b) as a function of transverse magnetic field ($G_1 = 3G_0$). A shift of position of V$_{th}$ peaks, denoted as V$_{pos}$. (b) Position of the highlighted V$_{th}$ peak in (a) (with respect to V$_2$). Solid traces show two types of fitting with (red) and without (green) $|B|^3$ term, respectively. Inset highlights contribution to peak shift from $B^2$ and $|B|^3$ terms. (c) Comparison between measured V$_{th}$ peak position and prediction based on $\frac{dlnG_2}{dV_2}$. Inset shows a zoom-in of low field regime.      }
	\label{fig:2}
	
\end{figure}   

After calibrating the measurement system at zero magnetic field, a transverse magnetic field $B_{\perp}$ was applied to the sample. $B_{\perp}$ is expected to influence the thermal property of the system because it affects the electron cooling and heat generation processes. At zero field, the temperature reaches maximal at
the center of the sample then decreases to reach the lattice
temperature at the contacts; in the presence of a transverse magnetic field, the heat is generated when the edge channels connect to the contacts, creating two hot spots close to each contact and located on opposite sides (top and bottom) of the sample\cite{KLA91}.

It is necessary to comment that the position of thermo-voltage peak with respect to gate voltage $V_i$ is mainly determined by the \textit{i}th QPC within the range of magnetic field in the present work. Using a single QPC or double QPCs leads to a similar result, as illustrated by Fig. S1.  On the other hand, the amplitude of the peak could be sensitive to the arrangement of the QPCs. For instance,   the amplitude of thermo-voltage peak with the QPCs locate at the same edge may differ from that with the QPCs locate at the opposite edges.

\begin{figure}
	
	\includegraphics[height=5.4in,width=3.2in]{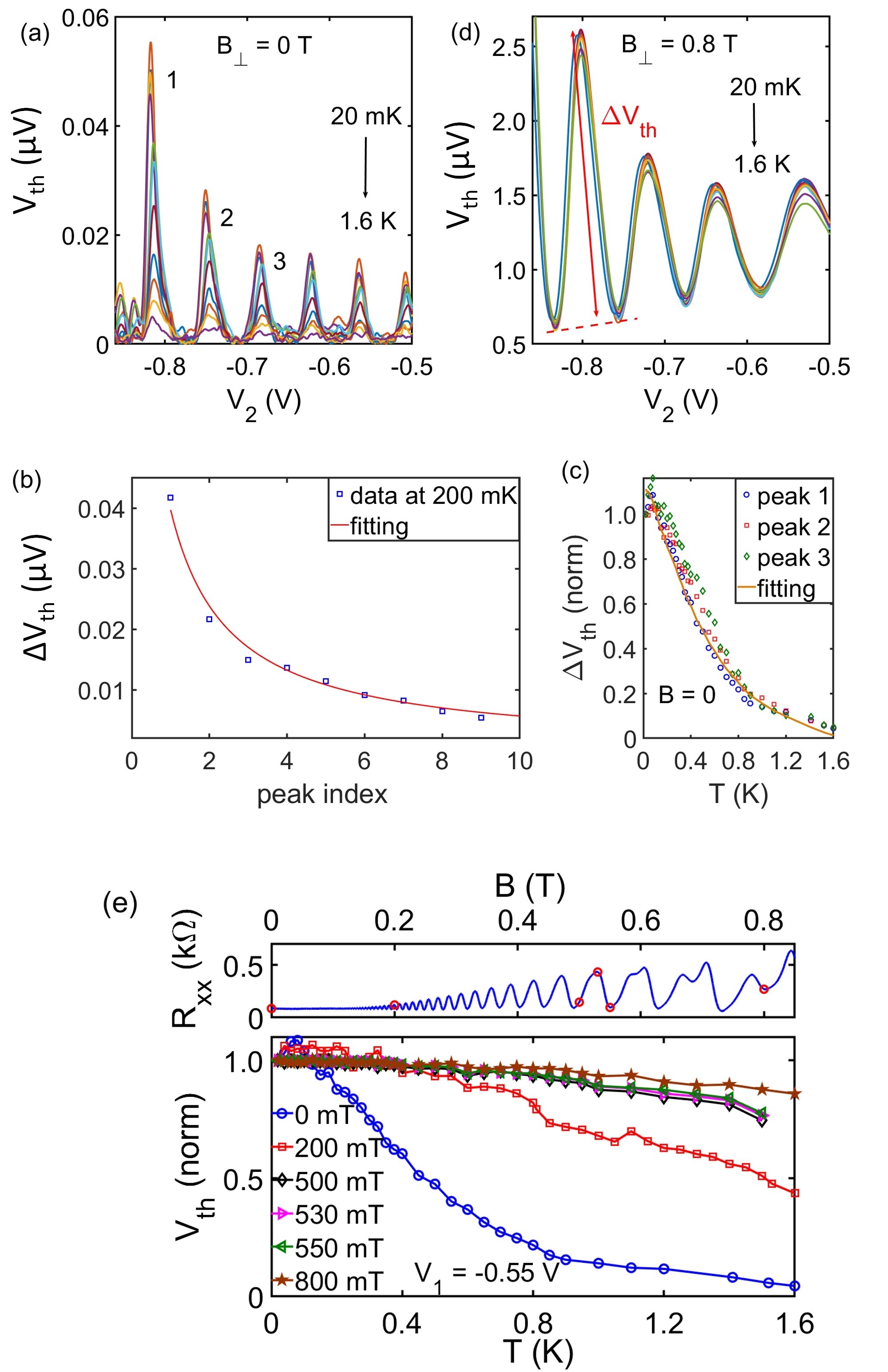}
	
	\caption{Lattice temperature dependence of V$_{th}$ peaks. (a) Temperature dependence at zero field. (b) Fitting of $\delta V_{th}$ against peak index at 200 mK at zero field. (c) Universal scaling behavior.(d) Temperature dependence at $B_{\perp}$ = 0.8 T. (e) Upper panel shows the SdH oscillation of the raw 2DEG. The red dots highlight the magnetic field values presented in the lower panel. The normalized magnitude of  V$_{th}$ peak 1, the highlighted peak in Fig. 2(a) [it occurs at different $V_2$ with increasing  $B_{\perp}$], at the different magnetic field as a function of lattice temperature ranging from 20 mK to 1.6 K. The measured results at higher temperature are normalized against that at 20 mK.     }  
	
	\label{fig:3}
	
\end{figure}  

Here we present a detailed magnetic field dependence of trace in Fig. \ref{fig:1}(b) (i.e. QPC1 is set to 3G$_0$ at zero field). It was found that the field had a twofold impact, as demonstrated in Fig. \ref{fig:2}$:$ first, the magnitude of $V_{th}$ peaks was enhanced noticeably by $B_{\perp}$, which is consistent with the predicted increase in thermopower (hence thermo-voltage) if size quantization owing to QPC is strong\cite{MS03}; this enhancement is also related to the fact the hot spot moves towards the edge of the sample; second and more important, a shift in the position of $V_{th}$ peaks was introduced by $B_{\perp}$ as shown in Fig. \ref{fig:2}(a) and (b) [also supplementary material, Fig.S2]. The position of $V_{th}$ peak changed rapidly when -3 T $< B_{\perp} <$ 3 T, and then saturated (when bulk Landau filling factor approaches $\nu$ = 2). The peak position follows such a relation $V_{pos} = c_3 |B|^3 + c_2 B^2 + c_0$ [red solid trace in Fig.3(b)]. The $B^2$ term dominated in small field regime whereas $|B|^3$ became the leading factor when $B_{\perp}  \geqslant$ 0.5 T.  In addition, it was also noticed that the $V_{th}$ peak position agreed well with that predicted by $\frac{\partial lnG_2}{\partial V_2}$ (i.e., Cutler-Mott relation) in small field ($B_{\perp} <$ 0.6 T); however, it diverged significantly from the prediction in a large field regime [Fig. \ref{fig:2} (c)]. Besides, $\frac{\partial lnG_2}{\partial V_2}$ split into two spin-resolved branches with increasing field due to Zeeman splitting whereas there was no sign of such splitting in $V_{th}$ peak.

In principle, both $V_{th}$ and $\frac{\partial lnG_2}{\partial V_2}$ monitor the position of the 1D subbands within the QPC, which determine the heat and charge transport through the QPC. The fact that $V_{th}$ peak correlates with $\frac{\partial lnG_2}{\partial V_2}$ in a small field regime where $B^2$ dominated, indicates that the $B^2$ term arises from magnetic depopulation induced 1D subbands rearrangement\cite{BTN86,VKW91}. The origin of the $|B|^3$ term, on the other hand, was unclear. Higher-order $B$ dependence can be induced if the electrostatic confinement is not parabolic, for instance, due to disorder\cite{BTN86}; however, the impact of non-parabolic confinement should also be revealed in electrical measurement $\frac{\partial lnG_2}{\partial V_2}$. Besides, the parabolicity of electrostatic confinement is double-checked by fitting the conductance to the saddle point model [note 2 and Fig. S3 of supplementary information]. Therefore the $|B|^3$ term is a unique characteristic of heat transport. 

Anomalous temperature dependence associated with the magnetic field was also observed, as shown in Fig. \ref{fig:3} [see note 3 of supplementary information for details on the fitting]. It was found that the magnitude of thermo-voltage peak attenuated rapidly with increasing lattice temperature at low magnetic field similar to the previous report\cite{ANS98,ANP00} as seen in Fig. \ref{fig:3}(a), and exhibited a universal scaling [Fig. \ref{fig:3}(b) and (c)]; on the other hand, it was rather insensitive to temperature change in the regime where $|B|^3$ term dominates, as exemplified by results at B = 0.8 T in Fig. \ref{fig:3}(d). It is intuitive to think that the observed temperature might be relevant with the edge channels because the QPC is attached to the hot spot near the Ohmic contact via the medium of edge channels. This scenario is somehow not supported by the data shown in Fig. \ref{fig:3}(e). It is particularly interesting to note $V_{th}$ obtained at 500 (Shubnikov$–$de Haas dip), 530 (peak) and 550 mT (dip) showed similar behavior, whereas it was well known that electrical measurements at SdH dip and peak had rather distinguishable temperature dependence\cite{WCT85,LCE99,ZTS05}.

The results at zero magnetic field can be well described by Cutler-Mott relation, which is a single-electron framework. Our results suggest this framework does not hold in the presence of a finite transverse magnetic field.  It is helpful to discuss several mechanisms beyond single-electron framework which are known to affect thermal conduction$:$

I. Phonon drag. Phonon drag augments thermopower (also thermo-voltage) with increasing magnetic field\cite{FDS88,ROJ88}. However, phonon drag usually freezes out below 0.6 K in the GaAs heterojunction\cite{ROJ88}. In our experiment, there was no sign of phonon drag switching on/off at the finite magnetic field.

II. A spin density wave (SDW). An SDW can be induced by a transverse magnetic field\cite{YL83,KSC86,NCY88,KFS96}, and it is separated from the single-particle mode by an energy gap\cite{PSD89,SGP94}. The melting temperature of an SDW is enhanced by the magnetic field\cite{NCY88, YCU90}, which might result in the observed evolution in temperature dependence. However, observations on an SDW in GaAs heterojunction usually involve excitation between several 2D subbands\cite{PSD89, SGP94}, which is unlikely to be found in the current experiment.  

III. Electron-electron (e-e) interaction. It is known that e-e interaction can make thermal conduction of a 1D system diverge from Cutler-Mott relation, such as the finite thermopower at '0.7 conductance anomaly'\cite{ANP00, BMF19}.  However, a detailed theory on QPC-mediated thermal conduction that incorporates electron-electron interaction, especially in the presence of a transverse magnetic field, is lacking. More specifically, it is difficult to comment at the moment why the e-e interaction prevents the occurrence of spin splitting in $V_{th}$.

In conclusion, we have observed unexplained thermal conduction through a 1D channel in the presence of a transverse magnetic field. The magnetic field and temperature dependence indicate that the thermal conduction at the finite magnetic field is beyond the single-particle picture. The results are important when employing the QPC to probe energy flow in integer/fractional quantum Hall or other exotic systems, where the magnetic field plays an important role. 

See the supplementary material for the results under different experimental conditions and details on the fitting of temperature dependence data.

The authors gratefully acknowledge fruitful discussions with Karl-Fredrik Berggren, Kalarikad Thomas, and James Nicholls.  The work was funded by United Kingdom Research and Innovation (UKRI).

\end{document}